\documentclass[12pt]{article}
\usepackage{amsmath,amssymb,amsfonts,amsthm,bm,slashed,graphics,graphicx}
\usepackage{epsfig,hyperref,subfigure,dsfont,multirow,psfrag,extarrows}
\usepackage{mathtools}
\usepackage[english]{babel}
\usepackage{bbold,float}
\usepackage{tikz}
\usepackage{mathtools}
\DeclarePairedDelimiter\ceil{\lceil}{\rceil}

\DeclarePairedDelimiter\norm{\lVert}{\rVert}
\theoremstyle{plain}

\newtheorem{lemma}{Lemma}
\newtheorem{theorem}{Theorem}
\newtheorem{proposition}{Proposition}
\newtheorem{definition}{Definition}
\newtheorem{corollary}{Corollary}
\newtheorem{remark}{Remark}

\theoremstyle{definition}
\DeclareMathOperator{\Tr}{Tr}

\newcommand{\al}{\alpha}

\newcommand{\Norm}{\Big\|}
\def\a{{\mathbb a}}
\def\C{{\mathbb C}}

\def\myI{{i}}
\newcommand{\prf}{{\noindent {\bf Proof}\quad }}

\newcommand{\lmbd}{\lambda}
\newcommand{\bea}{\begin{eqnarray}}
\newcommand{\ea}{\end{eqnarray}}
\newcommand{\bee}{\begin{equation}}
\newcommand{\ee}{\end{equation}}
\newcommand{\cA}{{\cal A}}
\newcommand{\cB}{{\cal B}}
\newcommand{\cZ}{{\cal Z}}
\newcommand{\cR}{{\cal R}}
\newcommand{\cD}{{\cal D}}
\newcommand{\cC}{{\cal C}}

\begin{document}

\title{Variational Loop Vertex Expansion\\ for Cumulants}
\author{V. Rivasseau\\ Universit\'e Paris-Saclay, CNRS/IN2P3\\ IJCLab, 91405 Orsay, France}
\date{} 
\maketitle
\begin{abstract}
Extending recent advances on constructive quantum field theory, we study cumulants of one of the simplest matrix models in the regime of bounded rank.
We analyze both ordinary cumulants and scalar cumulants, which arise from the Weingarten calculus and are essential to the topological expansion in quantum field theory.
Our results are valid for arbitrarily large positive coupling and provide new instances where 
techniques such as the variational approach are applied to cumulants. 
\end{abstract}

\noindent\textbf{keywords}
Cumulants; Constructive Field Theory; Matrix Field Theory; Variational Perturbation Theory.

\section{Introduction}

This work originates from the constructive field theory method known as the Loop Vertex Expansion (LVE) \cite{LVE}. 
The LVE is a constructive approach for quartic matrix models. It is designed to provide bounds that are uniform with respect to the matrix size. In its original formulation, the LVE combines an intermediate field representation with replica fields and a forest formula. This expresses the theory's free energy as a convergent sum over trees. 

Cumulants are statistical measures that provide insights into the shape and dependencies within a probability distribution. From a quantum field theory perspective, cumulants are essentially connected Schwinger functions.  In the context of quartic matrix models, cumulants play a crucial role in understanding the underlying structure and behavior of the system at hand.

It incorporates concepts from variational theory as exemplified by \cite{Saz}. 

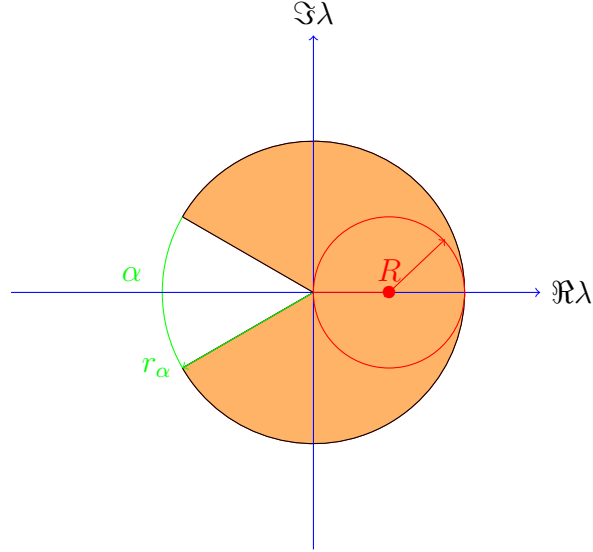
\begin{figure}[ht]
\begin{center}
\begin{tikzpicture}[scale=2]
\tikzstyle{red2}=[thick,  fill=red!40]
\draw[red] (-6.866,-.5) arc (-150:150:1) ;
\draw[red] (-6.866,-.5) -- (-6,0);
\draw[red] (-6.866,.5) -- (-6,0);
\filldraw[fill=orange!60] (-6.866,-.5) arc (-150:150:1)  --  (-6,0) -- cycle;
\draw[green] (-6.866,-.5)  arc (-150:-210:1) ;
\draw[green] (-7.2,0) node[above] {$\alpha$};
\draw[green,->]   (-6,0) -- (-6.866,-.5); 
\draw[green] (-6.866,-.5) node[left] {$r_\alpha$};
\draw[blue,->] (-8,0) -- (-4.5,0);
\draw (-4.5,0) node[right] {${\Re \lambda}$};
\draw [blue,->] (-6,-1.7) -- (-6,1.7);
\draw (-6,1.7) node[above] {${\Im \lambda}$}; 
\draw[red] (-5.5,0) circle (.5) ;
\draw[red,->] (-6,0) -- (-5.5,0);
\draw[red] (-5.5,0) node {$\bullet$} ;
\draw[red] (-5.5,0) node[above]{$R$} ;
\draw[red,->] (-5.5,0) -- (-5.13,0.35);
\end{tikzpicture}
\end{center}
\caption{This picture represents the LVE pacman domain of analyticity (in orange), with its angle $\al$ and radius $r_\al$ in green. 
In red the connection with Borel summability, with its circle ${\cal D}_{R}$ (see the Appendix).}\label{F1}
\end{figure}

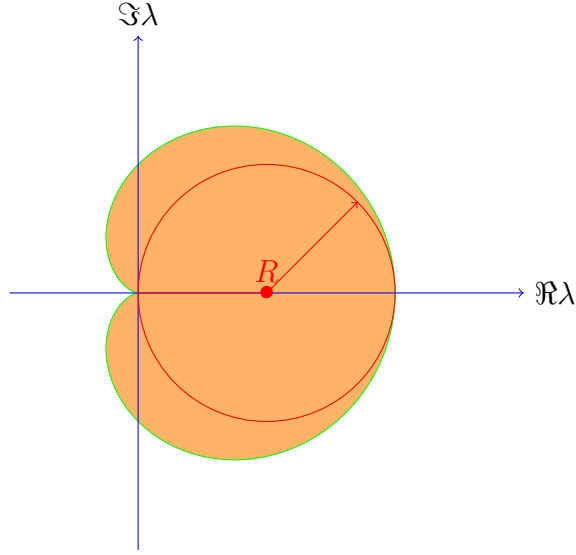
\begin{figure}[ht]
\begin{center}
\begin{tikzpicture}[scale=1.7]
\filldraw [draw=green,fill=orange!60,domain=-180:180,samples=400,scale=.5,variable=\t] plot (\t:  2+ 2*cos \t);
\draw[blue,->] (-1,0) -- (3,0);
\draw (3,0) node[right] {${\Re \lambda}$};
\draw [blue,->] (0,-2) -- (0,2);
\draw (0,2) node[above] {${\Im \lambda}$}; 
\draw[red] (1,0) circle (1) ;
\draw[red] (1,0) node {$\bullet$} ;
\draw[red] (1,0) node[above]{$R$} ;
\draw[red,->] (0,0) -- (1,0);
\draw[red,->] (1,0) -- (1.7071,0.7071);
\end{tikzpicture}
\end{center}
\caption{The cardioid domain of analyticity, also in orange. It can be thought of
as a union of pacman domains with the radius $r_\al$ shrinking with the size 
of the angle $\al$. 
In red the connection with Borel summability, with its circle ${\cal D}_{R}$.}\label{F2}
\end{figure}

\medskip
The standard LVE tools allow one to prove the analyticity of the free energy of the model in the pacman-like   (see Fig. \ref{F1})
or cardioid-like domains (see Fig. \ref{F2}) \cite{RivTenTr3,KRS,KRS1}. 

\medskip
The results of this study have significant implications for the broader field of combinatorial field theories involving random matrices and tensors. In summary, combining the LVE with variational perturbation theory provides a robust framework for analyzing the cumulants of quartic matrix models. This work extends the applicability of the LVE and opens new avenues for research in constructive field theory and its applications.

\section{Quick Summary  of 
\texorpdfstring{\cite{GurKra}}{GurKra}}
\label{sect2}

\medskip
We adopt the notations of \cite{Saz,GurKra,Riv1} with two exceptions, namely 

\begin{itemize}
\item  the parameter $a$ of \cite{Saz} is changed in $\a$ in our paper in order to distinguish it from the $J_{a,b}$ of \cite{GurKra},
\item the parameter $k$ is changed in ${\mathcal K}$ in our paper in all that is concerned with the cumulants to be in line with \cite{Riv1}.
\end{itemize}

The first step is to add the sources because \cite{Saz} does not have sources.
\bea
\label{Fundeq1}
{\cal Z}[\lambda,N;J,J^+]&=&
\frac{1}{\cZ[\lmbd,N]}\int dM\exp\big\{\sqrt{N}[\Tr(JM^{\dagger})+\Tr(MJ^{\dagger})]\big\}\nonumber\\
&&\exp \Big\{-\Tr(MM^{\dagger})-\frac{\lambda}{2N}\Tr(MM^{\dagger}MM^{\dagger})\Big\}.
\ea
The coupling constant is $\lambda= \rho e^{i \phi}$.
The partition function of \cite{Saz} is defined by
\bea\label{EQ2}
{\cal Z}[\lambda,N] =
\int dM\exp\Big\{-\Tr(MM^{\dagger})-\frac{\lambda}{2N}\Tr([MM^{\dagger}]^2)\Big\}\, 
\label{eq1}
\ea
where $M$ are complex $N\times N$ matrices, and the measure $dM$ is given by
\bea\label{EQ3}
dM=\pi^{-N^2}\prod_{1\leq i,j\leq N}d\Re( M_{ij})d\Im(M_{ij}) \, .
\ea

The free energy  without sources is 
\bea
F[\lambda, N] = -\frac{1}{N^2} \log {\cal Z}[\lambda, N]\,.
\label{eq2}
\ea

\medskip
The sharpest results with respect to the cumulants of our model were obtained in \cite{GurKra}. 
We work in the footsteps of \cite{GurKra}, an essential work for those interested in the subject of cumulants from a constructive perspective.
First, we extend the LVE to cumulants with the same definition as in \cite{GurKra}:  
\begin{definition}
\label{cumulants}
The \emph{cumulant of order ${\mathcal K}$} is 
\bea 
\mathfrak{K}^{{\mathcal K}}_{a_{1}b_{1}c_{1}d_{1},\dots,a_{{\mathcal K}}b_{{\mathcal K}}c_{{\mathcal K}}d_{{\mathcal K}}}(\lambda,N)=
\frac{\partial^{2}}{\partial J^{+}_{a_{1}b_{1}}\partial J^{}_{c_{1}d_{1}}}\cdots
\frac{\partial^{2}}{\partial J^{+}_{a_{{\mathcal K}}b_{{\mathcal K}}}\partial J^{}_{c_{{\mathcal K}}d_{{\mathcal K}}}}
\log{ {\cal Z}(J)} \bigg|_{\{J\}=0}\nonumber\\
=\sum_{\pi\in\Pi_{{\mathcal K}} }\mathfrak{K}^{\mathcal K}_{\pi}(\lambda,N)
\sum_{\rho,\sigma\in\mathfrak{S}_{{\mathcal K}}}
\prod_{1\leq l\leq {\mathcal K}}\delta_{c_{l},a_{\rho\tau_{\pi}\sigma^{-1}(l)}}\delta_{d_{l},b_{\rho\xi_{\pi}\sigma^{-1}(l)}} ,
\label{cum1}
\ea
where 
\begin{itemize}
\item
 $J_{ab}^{+}$ is the complex conjugate of $J_{ab}$, so that $J^{\dagger}_{ab}=J^{+}_{ba}$,
 \item the sums in \eqref{cum1} is over  partitions of ${\mathcal K}$ and over two permutations $\rho$ and $\sigma$ of ${\mathcal K}$ elements,
 \item $\mathfrak{K}^{\mathcal K}_{\pi}(\lambda,N)$ are called {\bf scalar cumulants},
 \item
$\tau_{\pi}$ and $\xi_{\pi}$ are arbitrary permutations such that $\tau_{\pi}(\xi_{\pi})^{-1}$ has a cycle structure corresponding to the partition $\pi$,
\item the normalization is chosen in such a way that the contribution of a genus $g$ graph with $b$ broken faces scales as $N^{2-2g-b}$, corresponding to the Euler characteristic of a surface with punctures.
\end{itemize}
\end{definition}

\begin{figure}[!t]
\begin{center}
\begin{tikzpicture}[scale=2]
\tikzstyle{red2}=[thick,  fill=red!40]
\draw (-0.5,-0.3) node {$O$} ;
\draw (2,0) node[above right]{$\frac14$} ;
\draw (0,1) node[above left]{$\frac18$} ;
\filldraw [draw=green,fill=orange!60,domain=-180:180,samples=400,scale=.5,variable=\t] plot (\t:  2+ 2*cos \t);
\draw (-0,0) node {$\bullet$} ;
\draw (2,0) node {$\bullet$} ;
\draw (0,1) node {$\bullet$} ;
\draw[->] (-2,0) -- (3,0);
\draw (3,0) node[right] {$x$};
\draw [->] (0,-2) -- (0,2.5);
\draw (0,2.5) node[above] {$y$}; 
\end{tikzpicture}
\end{center}
\caption{Open cardioid $\cC$ pictured in orange, frontier pictured in green. 
}
\label{cardio}
\end{figure}
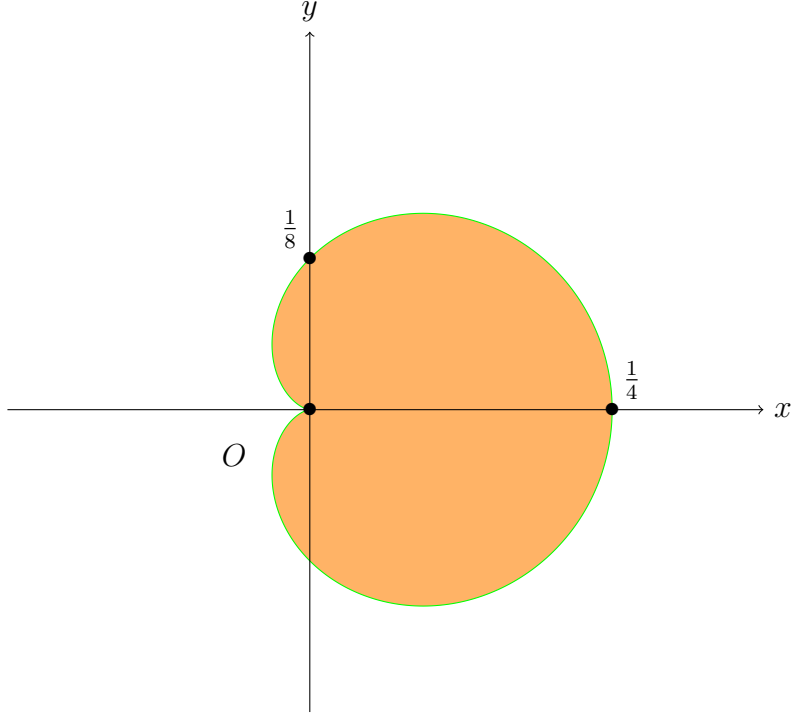

It was proven in \cite{GurKra} that 
the cumulants $\mathfrak{K}^{{\mathcal K}}_{a_{1}b_{1}c_{1}d_{1},\dots,a_{{\mathcal K}}b_{{\mathcal K}}c_{{\mathcal K}}d_{{\mathcal K}}}(\lambda,N)$ are analytic in the cardioid domain 
of the coupling constant $\lambda=\rho e^{i \phi}$  defined by (see Figure \ref{cardio})
\begin{equation}\label{eq6.0}
 \cC = \Big\{ \lambda \in \mathbb{C} \,\Big| \; 4 \rho < \cos^{2} \Big(\frac{\phi}{2}\Big) \Big\} \, .
\end{equation}
For the cardioid $\cC$ in question, see Figure \ref{cardio}.

\medskip
On the other part, the \emph{scalar cumulants} $\mathfrak{K}^{{\mathcal K}}_{\pi}(\lambda,N)$
can be written in terms of the Weingarten functions $\text{Wg}(\tau\sigma^{-1},N)$ \cite{Collins,ColSni}.
Denoting $U_{ab}^*$ the complex conjugate of $U_{ab}$ we have \cite{Collins}:
\begin{multline}
\int dU \; U_{a_{1}b_{1}}\dots U_{a_{{\mathcal K}}b_{{\mathcal K}}} U^{*}_{c_{1}d_{1}}\dots U^{*}_{c_{l}d_{l}}
=\\ \delta_{{\mathcal K}l}\sum_{\sigma,\tau\in \mathfrak{S}_{{\mathcal K}}} \delta_{a_{\tau(1)c_{1}}}\dots \delta_{a_{\tau({\mathcal K})}c_{{\mathcal K}}}
\delta_{b_{\sigma(1)}d_{1}}\dots \delta_{b_{\sigma({\mathcal K})}d_{{\mathcal K}}} \text{Wg}(\tau\sigma^{-1},N) \; ,
\label{Weingartenrelation}
\end{multline}
and the functions $\text{Wg}(\sigma,N)$ only depends on the cycle structure of $\sigma$. 
Furthermore the scalar cumulants $ \mathfrak{K}^{\mathcal K}_{\pi}(\lambda,N) $ are given by the expansion
\begin{equation}
\mathfrak{K}^{\mathcal K}_{\pi}(\lambda,N)=\sum_{T \text{ LVE tree with ${\mathcal K}$ cilia}}{\cal A}^{\pi}_{T}(\lambda,N)\label{Kpitree} ,
\end{equation}
where the cilia and ${\cal A}^{\pi}_{T}(\lambda,N)$ are defined in \cite{GurKra}.

\medskip
We associate to every LVE graph $(G,T)$ the \emph{amplitude} ${\cal A}_{G,T}[J,J^{\dagger},\lambda,N]$
defined as a Gaussian integral over $V(G)$ Hermitian matrices $(A_{i})_{1\leq i\leq V(G)}$ (each one of size $N\times N$):
\bea
{\cal A}_{G,T}[J,J^{\dagger};\lambda,N]&=&\frac{(-\lambda)^{\vert E(G)\vert }N^{\vert V(G)\vert -\vert E(G)\vert }}{\vert V(G)\vert !}
\mathop{\int}\limits_{1\geq s_{1}\geq\cdots\geq s_{L(G,T)}\geq 0}\,\prod_{e\in L(G,T)}ds_{e} \nonumber\\
&&\times \mathop{\int}\limits_{[0,1]} \prod_{e\in E(T)} dt_{e}
  \left( \prod_{e=(i,j)\in L(G,T)}\mathop{\text{inf}}\limits_{e'\in P_{i\leftrightarrow j}^{T}} t_{e'}  \right) \\ \nonumber
&& \hskip-2.5cm\times \int d\mu_{s_{\vert L(G,T)\vert }C_{T}}(A)\prod_{f\in F(G)}\Tr\bigg\{\mathop{\prod}\limits_{c\in\partial f}^{\longrightarrow}
\bigg(1-\text{i}\sqrt{\frac{\lambda}{N}}\,A_{i_{c}}\bigg)^{-1}(JJ^{\dagger})^{\eta_{c}}\bigg\} \; ,\label{LVEamplitude}
\ea
where:
\begin{itemize}
\item ${\displaystyle\mathop{\prod}\limits_{c\in\partial f}^{\longrightarrow}}$ is the oriented product around the corners $c$ 
on the boundary $\partial f$ of the face $f$.
\item $i_{c}$ is the label of the vertex the corner  $c$ belongs to.
\item $\eta_{c}=1,0$ depending on whether $c$ is followed by a cilium (1) or not (0).
\end{itemize}
Then these theorems, corollary and propositions are also proven in \cite{GurKra}:

\begin{proposition}\label{prop:ampliWeingarten}
The amplitude of a LVE graph $G$
in \eqref{LVEamplitude}  expands in trace invariants as:
\bea \label{eq10}
{\cal A}_{G,T}[J,J^{\dagger},\lambda,N]  &=&
\sum_{\pi\in\Pi_{k}}{A}_{G,T}^{\pi}(\lambda,N) \; \Tr_{\pi}(JJ^{\dagger}) \; ,
\ea
with
\bea\label{eq11}
{A}_{G,T}^{\pi}(\lambda,N)&=&\frac{(-\lambda)^{\vert E(G)\vert }N^{\vert V(G)\vert -\vert E(G)\vert }}{\vert V(G)\vert !}
\mathop{\int}\limits_{1\geq s_{1}\geq\cdots\geq s_{\vert L(G,T)\vert }\geq 0}\,\prod_{e\in L(G,T)}ds_{e}\nonumber\\
&& \hskip-2cm \times\int \prod_{e\in E(T)} dt_{e}
 \left( \prod_{e=(i,j) \in L(G,T)}\mathop{\text{inf}}\limits_{e'\in P_{i\leftrightarrow j}^{T}} t_{e'} \right) 
\int d\mu_{s_{\vert L(G,T)\vert }C_{T}}(A) \\
\nonumber&&\hskip-3cm\times\sum_{\substack{\tau,\sigma\in{\mathfrak S}_{k}\\C(\sigma)=\pi}}
\sum_{1\leq p_{1},\dots,p_{k}\leq N}\text{Wg}(\tau\sigma^{-1},N)
\prod_{1\leq m\leq F(G)-B(G)}\Tr\Big[Y^{m}\Big]\prod_{1\leq l\leq k}
X^{l}_{p_{\tau(l) }p_{\zeta(l)}}.
\ea
where $X$ and $Y$ are defined in \cite{GurKra} as the product of the some resolvents, respectively 
located on the corners separating the cilia labeled $l$ and $\zeta(l)$, and around the unbroken face labeled $m$. 
\end{proposition}

Then we quote an equation of \cite{GurKra} 
\bea\label{eq12}
\prod_{1\leq m\leq b}\Tr\Big[JJ^{\dagger}
\mathop{\prod}\limits_{1\leq r\leq {\mathcal K}_{m}}^{\longrightarrow}X^{i^{m}_{r}}\Big]=
\sum_{1\leq p_{1}, q_1 \dots \leq N} \prod_{1\leq l\leq {\mathcal K}} (JJ^{\dagger})_{p_{l}q_{l}}  X^{l}_{q_{l}p_{\zeta(l)}} \;.
\ea

If the LVE graph $(G,T)$ is reduced to a tree we use the shorthand notation ${A}_{T}^{\pi}(\lambda,N) $ instead of 
${A}_{(T,T)}^{\pi}(\lambda,N)$. 

\begin{proposition}[Scalar cumulants]
\label{structure:prop}
The order ${\mathcal K}$ cumulants can be written as a sum over  partitions of ${\mathcal  K}$ and over two permutations of ${\mathcal  K}$ elements:
\bea
\label{structure:eq}
{{{\mathfrak K}}}^{\mathcal K}_{a_{1}b_{1}c_{1}d_{1},\dots,a_{^{\mathcal K}}b_{^{\mathcal K}}c_{^{\mathcal K}}d_{^{\mathcal K}}}(\lambda,N)&=&\!\sum_{\pi\in\Pi_{^{\mathcal K}} }{\mathfrak K}^{\mathcal K}_{\pi}(\lambda,N)
 \\&& \times
 \sum_{\rho,\sigma\in\mathfrak{S}_{{\mathcal  K}}} \hskip-.1cm
\prod_{1\leq l\leq {\mathcal  K}}\delta_{c_{l},a_{\rho\tau_{\pi}\sigma^{-1}(l)}}\delta_{d_{l},b_{\rho\xi_{\pi}\sigma^{-1}(l)}} ,\nonumber
\ea
where $\tau_{\pi}$ and $\xi_{\pi}$ are arbitrary permutations such that $\tau_{\pi}(\xi_{\pi})^{-1}$ has a cycle structure 
corresponding to the partition $\pi$ and the \emph{scalar cumulants} $ {\mathfrak K}^{\mathcal K}_{\pi}(\lambda,N) $ 
are given by the expansion:
\bea
\label{structure:eq1}
{\mathfrak K}^{\mathcal K}_{\pi}(\lambda,N)&=&\sum_{T \text{ LVE tree with ${\mathcal  K}$ cilia}}{\cal A}^{\pi}_{T}(\lambda,N)\label{Kpitree1} \; .
\ea
\end{proposition}

\begin{theorem}[Analyticity and bound for scalar cumulants] \label{treecumulants:thm}
The series:
\bea\label{eq15}
{\mathfrak K}^{\mathcal K}_{\pi}(\lambda, N)&=&\sum_{T\text{ LVE tree with ${\mathcal  K}$ cilia}}{\cal A}^{\pi}_{T}(\lambda, N)\;,
\ea
defines an analytic function of $\lambda\in{\cal C}$. Moreover, each term in this sum is bounded (for $N$ large enough) as:
\bea\label{treecumulantsbound}
\big|{\cal A}^{\pi}_{T}(\lambda, N)\big|\leq\frac{N^{2-|\pi|}|\lambda|^{\vert E(T)\vert }\,({\mathcal  K}!)^2 \, 2^{2{\mathcal  K}}}{(\cos\frac{\arg\lambda}{2})^{2\vert E(T)\vert +{\mathcal  K}}\,\vert V(T)\vert !}\; ,
\ea
where $|\pi|$ is the number of integers in the partition $\pi$ of ${\mathcal  K}$ (number of cilia).
\end{theorem}

\begin{figure}[!htb]
\centering
\begin{tikzpicture}[>=latex,scale=0.5]
\filldraw[draw=black,fill=orange!40] plot (14.7,1) arc (20: 161: 7.6) -- (0,0) -- (15,0) -- cycle;
\draw[color=black,fill=blue!20, very thick](0,0) circle (1);  
\draw[color=black,fill=blue!20, very thick](5,0) circle (1); 
\draw[color=black,fill=blue!20, very thick](10,0) circle (1); 
\draw[color=black,fill=blue!20, very thick](15,0) circle (1); 
\draw[color=black,fill=blue!20, very thick](5,-5) circle (1); 
\draw (0,0) node{$1$} ;\draw (5,0) node{$2$} ;\draw (10,0) node{$3$} ;\draw (15,0) node{$4$} ;\draw (5,-5) node{$5$} ;
\draw (7.5,7.7) node{$\text{loop 1}$} ;\draw (12,-5) node[below]{$\text{loop 2}$} ;\draw (5,2) node[above]{$\text{cilium 1}$} ;\draw (10,2) node[above]{$\text{cilium 2}$} ;
\draw[color=red,line width=4pt] (1,0) -- (4,0)  ;
\draw[color=red,line width=4pt] (6,0) -- (9,0)  ;
\draw[color=red,line width=4pt] (11,0) -- (14,0) ;
\draw[color=red,line width=4pt] (5,-1) -- (5,-4) ;
\draw[color=green, line width=4pt] (5,1) -- (5,2) ;
\draw[color=green, line width=4pt] (10,1) -- (10,2) ;
\draw[color=violet,line width=4pt] (14.7,1) arc (20: 161: 7.6); 
\draw[color=violet,line width=4pt]  (6,-5) arc (-90: -39: 11); 
\end{tikzpicture}
\caption{A LVE graph with five vertices colored in blue, four propagators colored in red, two loops colored in violet, two cilia colored in green and one broken face colored in orange.} 
\label{Fig3a}
\end{figure}
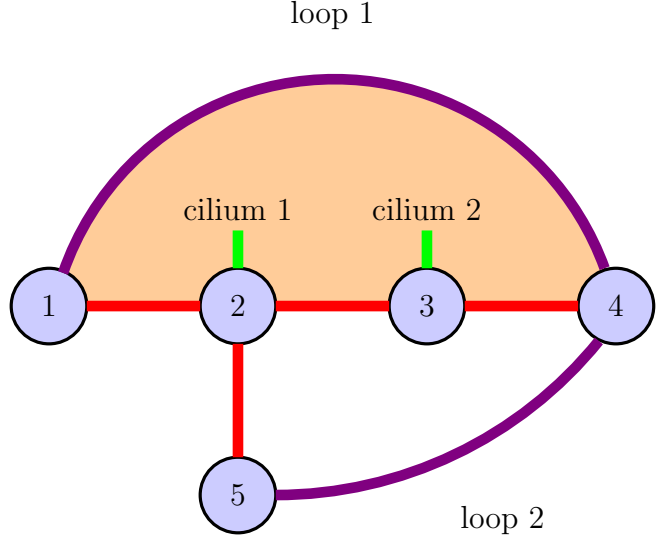

\begin{theorem}[Perturbative expansion with remainder] \label{perturbativecumulants:thm}
The perturbative expansion of the cumulants reads:
\bee\label{eq17}
{\mathfrak K}^{\mathcal K}_{\pi}(\lambda, N)=\hskip-.1cm\sum_{\substack{G\text{ ribbon graph with ${\mathcal  K}$ cilia,}\\
\text{broken faces corresponding to $\pi$ and $\vert E(G)\vert \leq n$}}}\hskip-.1cm
\frac{(-\lambda)^{\vert E(G)\vert }N^{\chi(G)}}{|\text{Aut}(G)|}+{\cal R}^{\mathcal K}_{\pi,n}(\lambda,N) .
\ee
The perturbative remainder  ${\cal R}^{\mathcal K}_{\pi,n}(\lambda,N)$ is a sum over LVE graphs with  ${\mathcal  K}$ cilia, at least $n+1$ edges and at most $n+1$ loop edges,
\bee\label{eq18}
{\cal R}^{\mathcal K}_{\pi,n}(\lambda,N)= \sum_{\substack{(G,T)\text{ LVE graphs with broken structure corresponding to $\pi$,}\\
\vert E(G)\vert \geq n+1 \text{ and } \vert L(G,T)\vert \leq n+1}}{\cal A}^{\pi}_{(G,T)}(\lambda, N) \; .
\ee
The perturbative reminder is analytic for $\lambda \in {\cal C}$ and for any $\lambda \in {\cal C}$ and $N$ large enough it obeys the bound:
\bea\label{eq19}
\Big|{\cal R}^{\mathcal K}_{\pi,n}(\lambda,N)\Big|
&\le& N^{2-|\pi|} \left( \frac{2^{3{\mathcal  K}-1}{\mathcal  K}!}{ \big(\cos\frac{\arg\lambda}{2}\big)^{{\mathcal  K}} } \right) (n+1)!   
\left(  \frac{ 4 |\lambda|}{  \big(\cos\frac{\arg\lambda}{2}\big)^{2}} \right)^{n+1} \nonumber
\\&&\hskip1cm \times
\left(\frac{    \frac{ 4 |\lambda|}{  \big(\cos\frac{\arg\lambda}{2}\big)^{2} }   }
{ \left( 1 -  \frac{ 4 |\lambda|}{  \big(\cos\frac{\arg\lambda}{2}\big)^{2} } \right)^{n+2} }
+ 2^{{\mathcal  K}+n+2 }\right)\; .
\ea
\end{theorem}

\begin{corollary}[Borel summability]
\label{Boreltheorem}
The rescaled cumulants  $N^{-2+|\pi|}{\mathfrak K}^{\mathcal K}_{\pi}(\lambda, N)$ (with $|\pi|$ the number of parts in the partition $\pi$) 
are Borel summable in $\lambda$ at the origin, uniformly in $N$, so that
\bee\label{eq20}
{\mathfrak K}^{\mathcal K}_{\pi}(\lambda, N)=\int_{0}^{\infty}\!\!\!\!ds\,\text{e}^{-\frac{s}{\lambda}}\bigg(\sum_{n\geq {\mathcal  K}}\frac{a_{\pi,n}(N)}{n!}s^{n}\bigg) \;, 
\ee
in a disc included in ${\cal C}$ tangent to the imaginary axis at the origin and independent of $N$.
\end{corollary}

\begin{theorem}[Topological expansion]
\label{topologicalcumulants:thm}
The cumulants ${\mathfrak K}^{\mathcal K}_{\pi}(\lambda, N)$ are expanded in inverse powers of $N$ as
\bee\label{eq21}
{\mathfrak K}^{\mathcal K}_{\pi}(\lambda, N)=\sum_{h=0}^{g} N^{2-2g-|\pi|}{\mathfrak K}^{\mathcal K}_{\pi,h}(\lambda)+\widetilde{R}_{\pi,g}(\lambda,N) \;, 
\ee
where ${\mathfrak K}^{\mathcal K}_{\pi,h}(\lambda)$ is a sum over ciliated ribbon graphs of genus $h$ whose broken faces correspond to the partition $\pi$, convergent for $|\lambda|<\frac{1}{12}$:
\bee\label{eq22}
{\mathfrak K}^{\mathcal K}_{\pi,h}(\lambda)=\sum_{\substack{G\text{ ribbon graph with genus $h$}\\
\text{and broken faces corresponding to $\pi$}}} \frac{(-\lambda)^{\vert E(G)\vert }}{\vert \text{Aut} \ G\vert } \; .
\ee
The topological remainder $\widetilde{R}^{\mathcal K}_{\pi,g}(\lambda,N)$ is a sum over LVE 
graphs with broken faces corresponding to $\pi$, genus $g+1$ and such that, if we remove the loop edge of highest label, we get a genus $g$ graph 
\bee\label{eq23}
\widetilde{{\cal R}}^{\mathcal K}_{\pi,g}(\lambda,N)=
\sum_{\substack{(G,T)\text{ LVE graphs with broken faces corresponding to $\pi$,}\\
g(G)=g+1 \text{ and } g(G-e_{\vert L(G,T)\vert })=g}}
{\cal A}^{\pi}_{(G,T)}(\lambda, N) \; .
\ee
This series converges for $\lambda\in\widetilde{\cal C}$ and in this domain the topological reminder is bounded by 
\bea\label{eq24}
 \big|\widetilde{R}^{\mathcal K}_{\pi,g}(\lambda,N)\big| &\leq& 
 N^{2-2(g+1)-|\pi|} \frac{2^{3{\mathcal  K}} {\mathcal  K}! }{ \big(\cos\frac{\arg\lambda}{2}\big)^{{\mathcal  K}} } C''_{g+1}
 \\ && \times
\left( \frac{12|\lambda|}{\big(\cos\frac{\arg\lambda}{2}\big)^{2} } \right)^{2g+2} 
\frac{(4g+{\mathcal  K}+1)! }{ \Big( 1 -  \frac{12|\lambda|}{\big(\cos\frac{\arg\lambda}{2}\big)^{2} } \Big)^{4g+{\mathcal  K}} } \;, \nonumber
\ea
with $C''_g$ a constant depending only on the genus.
\end{theorem}
 
\medskip
We conclude this section by an other  lemma proved in \cite{GurKra}:
\begin{lemma}
\label{lemm1}
The number of LVE trees with $n$ edges and  ${\mathcal K}$ cilia is
\bea\label{eq27}
{\cal N}(n,{\mathcal K})&=&\frac{(2n+{\mathcal K}-1)!\,(n+1)!}{(n+{\mathcal K})!\,(n+1-{\mathcal K})!\,{\mathcal K}!}\\
&\leq& 2^{2n+{\mathcal K}-1}\,(n-1)!\,\frac{(n+1)!}{(n+1-{\mathcal K})!\,{\mathcal K}!} \; \\
&\leq&    2^{3n+{\mathcal K}}\ (n-1)!
\ea
\end{lemma}
\section{Useful Lemmas}
\label{sect3}

Let us introduce the intermediate representation which is the core of the LVE \cite{LVE}. 
The partition function of our model is, in this representation: 
\bea
&&\hskip-1cm\cZ[\lmbd,N]=\int dM \int dA\, \exp \Big\{-\frac{1}{2}\Tr(A^{2})
+ \myI \Tr\big(A \big[\sqrt{\frac{\lmbd}{N}}MM^{\dagger} \big)\big]\Big\}\,,
\label{eq3a}
\ea
where $M$ are complex $N\times N$ matrices, the $A$ field is realized by the $N\times N$ Hermitian matrices, and the integral over $A$ is assumed to be normalized, so
\bee\label{eq29}\int dA \exp\Big\{-\frac{1}{2} \Tr(A^2)\Big\} = 1\,.
\ee

\medskip
In \cite{Saz} the author changes  the initial approximation from $\Tr(MM^\dagger)$ to $\a
\Tr(MM^\dagger)$ (which correspond to a mass renormalisation).  Thus in \cite{Saz} the normalization of the partition function can be also written as
\bea
\cZ[\lmbd,N]&=& K[\lmbd, N, \a] \int dA\, \exp\bigg\{ - \frac{1}{2}\Tr(A^{2})-N{\cal S}[\lmbd, N, \a](A) \bigg\}, \label{eq4}
\ea
while $ K[\lmbd, N, \a]$ is defined as
\bea\label{eq31}
K[\lmbd, N, \a] &=& \a^{-N^2}\exp\Big\{\frac{N^2 (1 - \a)^2 }{2 \lmbd}\Big\}\,.
\ea

Hereafter, we write the parameter $\a$ describing the initial approximation as:
\bea
\a &=& x \sqrt{\lmbd} e^{\myI\psi}\,, \qquad -\frac{\pi}{2} < \psi + \frac{\phi}{2} < \frac{\pi}{2}\,, \qquad x > 0\,.
\label{arep}
\ea
We cite a useful lemma from \cite{Saz,GurKra}
 \begin{lemma}
 \label{boundresolvent}
According to the notations above, we have $\frac{\lmbd}{\a^2}=x\mathrm{e}^{-2\myI\psi}$ with $x>0$, then for $\psi \in (-\pi/2, \pi/2)$:
 \bea\label{eq33}
 \Norm\Big(1-\frac{\myI}{\a}\frac{\sqrt{\lmbd}}{\sqrt{N}}A\Big)^{-1}\Norm\leq\frac{1}{\cos \psi} \, ,
 \ea
 where $ \Norm \cdot \Norm $ stands for the operator norm\footnote{ In the rest of our paper the notation $\Vert ... \Vert$ is always referred to the operator  norm.}.
 \end{lemma}

In a simplified writing we use $ {\cal Z}(J)$ to mean the full 
${\cal Z}[\lambda,N;J,J^+]$.
Therefore, introducing the intermediate field representation like in \cite{Saz}, we obtain for the cumulants defined in \eqref{Fundeq1}
\begin{eqnarray}
{\cal Z}(J)&=&\frac{1}{\cZ[\lmbd,N]}
\int dM \exp \big\{\sqrt{N}[\Tr(JM^{\dagger})+\Tr(MJ^{\dagger})]\big\} \nonumber\\&& 
\times\int dA\, \exp \Big\{-\frac{1}{2}\Tr(A^{2})
- \a \Tr(MM^{\dagger})\nonumber\\&&\hskip-2.5cm + \myI \Tr\Big(A \Big[\sqrt{\frac{\lmbd}{N}}MM^{\dagger}+ \frac{(1 - \a) \sqrt{N}}{\sqrt{\lmbd}}\mathbb{1}\Big]\Big) + \Tr\big[\frac{(1 - \a)^2N}{2\lmbd}\mathbb{1}\big]\Big\}\,.
\label{eq3}
\end{eqnarray}

The integral over the initial degrees of freedom, the matrices $M$ and $M^{\dagger}$, is Gaussian with a linear term, 
therefore it can be evaluated. First let us rewrite the equation \eqref{eq3} as
\bea 
{\cal Z}(J)&=&\frac{1}{\cZ[\lmbd,N]}
\int dM e^{-\Tr \big( \a  MM^{\dagger} -\myI A\sqrt{\frac{\lmbd}{N}}MM^{\dagger}  
-\sqrt{N}[JM^{\dagger}+MJ^{\dagger}] \big)}\nonumber\\&&
\times \int dA\, e^{ -\frac{1}{2}\Tr A^{2}
+ \myI A \Tr\big[\frac{(1 - \a) \sqrt{N}}{\sqrt{\lmbd}}\mathbb{1}\big] 
+ \Tr\big[\frac{(1 - \a)^2N}{2\lmbd}\mathbb{1}\big]}\,.
\label{fundeq2}
\ea

The corresponding covariance matrix in  the matrices $M$ and $M^{\dagger}$ is given by \cite{Saz,Guraubook}
\bea
\label{eq32}
C&=&  \big[ \a -\myI   \frac{\sqrt{\lmbd}}{\sqrt{N}}   A) \big] \otimes \mathbb{1} ,
\ea
and the determinant is
\bea
\det \left[ \Big(\a-\myI    \frac{\sqrt{\lmbd}}{\sqrt{N}}    A \Big)\otimes \mathbb{1} \right] 
=\a^{N^2} \exp \bigg\{N\Tr\log\Big(\mathbb{1}-\frac{\myI}{\a}  \frac{\sqrt{\lmbd}}{\sqrt{N}}  A\Big)\bigg\} \, .
\ea

\begin{remark}
Please note that the right and left sides of the matrices $M$, $M^{\dagger}$, $J$ and $J^{\dagger}$ are not interchangeable.
\end{remark}

\medskip
Finally we arrived at the analog of equation (3.7) of \cite{Saz}, but this time \emph{with sources}:
\begin{lemma}\label{lemma1}
The following equation holds: 
\bea
\label{fundeq6.bis}
\cZ(J)\!&\!=\!&\!K[\lambda, N, \a]\! \int \!dA e^{-\frac{1}{2}\Tr A^{2}-N{\cal S}[\lambda, N, \a](A)
-N\Tr ( J^{\dagger} C J )    } .
\ea
\end{lemma}
\prf
Using \eqref{eq32} and Appendix B.1.2 of \cite{Guraubook} we arrived at
\bea
\label{eq39.0}
\int dM e^{- \Tr \big( M C^{-1} M -  \sqrt{N}[JM^{\dagger}+MJ^{\dagger}]\big) }&=& e^{-N \Tr (J^{\dagger} C J)}.
\ea
Next we rewrite \eqref{fundeq2} as
\bea 
\label{fundeq3}
\cZ(J)&=&
\frac{1}{\cZ[\lmbd,N]}
 \int dA\, e^{ -\frac{1}{2}\Tr A^{2}  - N \Tr ( J^{\dagger} C J ) 
+ \myI A \Tr\big[\frac{(1 - \a) \sqrt{N}}{\sqrt{\lmbd}}\mathbb{1}\big] 
 + \Tr\big[\frac{(1 - \a)^2N}{2\lmbd}\mathbb{1} \big]}\nonumber\\
& =&K[\lambda, N, \a]\! \int \!dA e^{-\frac{1}{2}\Tr A^{2}-N{\cal S}[\lambda, N, \a](A)
-N\Tr ( J^{\dagger} C J )    }.
\ea
Hence the lemma is proved.
\qed

\begin{definition}
Imitating \cite{GurKra} but taken into account \cite{Saz} we define $\cR$ and $\cD$ as two  $N$ by $N$ matrices:
\bea
\label{eq38}
\cR &:=& C^{-1} =\Big([\a -\myI   \frac{\sqrt{\lmbd}}{\sqrt{N}}  A]\otimes\mathbb{1}\Big)^{-1}.
\ea
\bea
\label{eq39}
\cD &:=& \Big[ \prod_{l'=1}^{\mathfrak K} \prod_{l''=1}^{\mathfrak K}\delta_{c_{l'}e_{l''}}\delta_{d_{l'}f_{l''}}\delta_{a_{l'}f_{l''}}\delta_{b_{l'}g_{l''}}  \Big].
\ea 
and to further simplify our notation let put
\bea
\label{eq40}
{\cal S}(A)&:=&{\cal S}[\lmbd, N, a](A),\\
\mathfrak{K}^{{\mathcal K}}_{\{abcd\}}&:=&
\mathfrak{K}^{{\mathcal K}}_{a_{1}b_{1}c_{1}d_{1},\dots,a_{{\mathcal K}}b_{{\mathcal K}}c_{{\mathcal K}}d_{{\mathcal K}}}(\lambda,N).
\label{eq41}
\ea
\end{definition}

\section{Main Results}
\label{sect4}

\medskip
\begin{definition}
\noindent
For $0<\epsilon\le \frac{\pi}{2}$,
${\cal E}\in \C$ is defined as a subset $
\lambda = \rho e^{i\phi}$, $\lambda \neq 0$,
and $\vert \phi \vert < \pi - \epsilon$.
\end{definition}
See ${\cal E}$ 
for $\epsilon = \frac{\pi}{4}$  represented in Figure \ref{CuE}. Remark that  ${\cal E}$ contais the disk $\cD_R$ of the Appendix and this for any $R$: thus the Nevanlinna-Sokal theorem is applicable to ${\cal E}$.

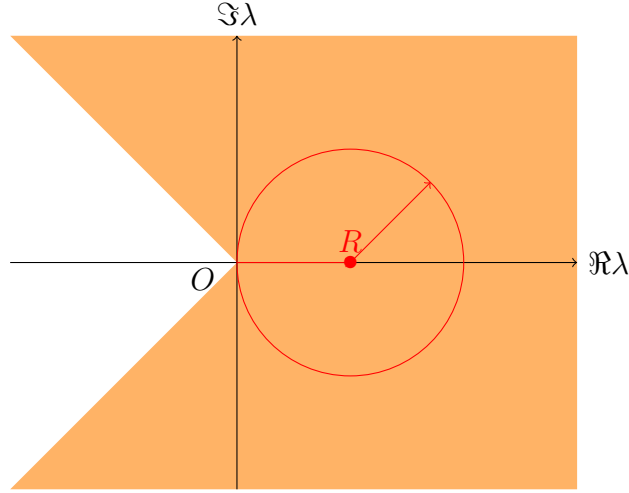
\begin{figure}[!h]
\begin{center}\begin{tikzpicture}[scale=1.5]
\draw (3,0) node[right] {$\Re \lambda$};
\draw (0,2) node[above] {$\Im \lambda$};
\fill[color=orange!60] (-2,-2) -- (3,-2) -- (3,2) -- (-2,2) -- (0,0) -- cycle ;
\draw (-0.3,-0.15) node {$O$} ;
\draw[->] (-2,0) -- (3,0);
\draw [->] (0,-2) -- (0,2);
\draw[red] (1,0) circle (1) ;
\draw[red] (1,0) node {$\bullet$} ;
\draw[red] (1,0) node[above]{$R$} ;
\draw[red,->] (0,0) -- (1,0);
\draw[red,->] (1,0) -- (1.7071,0.7071);
\end{tikzpicture}
\end{center}
\caption{${\cal E}$ for $\vert \phi \vert < \frac{3\pi}{4}$, $\epsilon = \frac{\pi}{4}$ pictured in orange.
In red the connection with Borel summability, with its circle ${\cal D}_{R}$. Any value of $R$ is acceptable.}
\label{CuE}
\end{figure}

\medskip
Let $\sum_{m=0}^{\infty}a_{m}(N,J)\lambda^m$ be the formal sum over $\mathfrak{K}^{{\mathcal K}}_{\{abcd\}}$.
We state our main result in the form of two theorems:
\begin{theorem}\label{mainTheorem1}
Let $1\le {\mathcal K}$ and the condition  $\norm{JJ^+} \le 1$ be fulfilled. 
The cumulants $\mathfrak{K}^{{\mathcal K}}_{\{abcd\}}$ are analytic in $\lambda \in {\cal E}$ uniformly in  $N$. 
The remainder at order $n$ obeys:
\bea
| {\cal R}^{\mathcal K}_{n}(\lambda,N,J)| :=  \big| \mathfrak{K}^{{\mathcal K}}_{\{abcd\}}
-\sum_{m=0}^{n}a_{m}(N,J)\lambda^{m} \big|\le C \sigma^{n+1}\, (n+1) ! \, | \lambda |^{n+1} .\label{BorLeRSum01}
\ea
with $C$ and $\sigma$ two positive constants that do not depend on $N$ and $J$.
Therefore it obeys the theorem of Nevanlinna-Sokal, stated in the Appendix  of this article
with $z \to \lambda$ and $\omega\to \big\{ N,J \big\}$.
\end{theorem}
\begin{remark}
This theorem is formulated in \cite{Riv1}, but for $\lambda \in \cC$ and under the condition ${\mathcal K} \le {\mathcal K}_{\max}$.
We add the extension to $\lambda \in {\cal E}$ and remove the restriction ${\mathcal K} \le {\mathcal K}_{\max}$.
But in \cite{Riv1} the analog of Theorem \ref{mainTheorem1} apply to a perturbation $(M M^\dagger )^p$ \emph{with $p\ge 2$.}
\end{remark}
\begin{theorem}
\label{mainTheorem2}
Under the same conditions, the expansion 
\bea\label{eq46}
{\mathfrak{K}}^{\mathcal K}_{\pi}(\lambda, N)=\sum_{T\text{ LVE tree with ${\mathcal K}$ cilia }}{\cal A}^{\pi}_{T}(\lambda, N)\;
\ea
is analytic in $\lmbd \in {\cal E}$ uniformly in $N$. Moreover, each term in this sum is bounded as:
\begin{equation}\label{treecumulantsbound1}
\big|{\cal A}^{\pi}_{T}(\lambda, N)\big|\leq\frac{N^{2-|\pi|}|\lambda|^{\vert e(T)\vert }\,({\mathcal K}!)^2 \, 2^{2{\mathcal K}}}
{(\cos\frac{\arg\lambda}{2})^{2\vert e(T)\vert +{\mathcal K}}\,\vert v(T)\vert !} \; ,
\end{equation}
where $|\pi|$ is the number of integers in the partition $\pi$. 
Furthemore it obeys the theorem stated in the Appendix of this article  (Nevanlinna-Sokal) with  $z \to \lambda$,
$\omega\to N$.
\end{theorem}
\begin{remark}
This theorem has already been formulated in \cite{GurKra}, \emph{but for $\lambda \in \cC$}. We add the extension to 
$\lambda \in {\cal E}$.
\end{remark}

\section{Lemmas preliminary to the two results}
\label{sect5}

It is essential to treat the part associated to  $\lambda \in {\cal E}$, hence make good use of \cite{Saz} 
since  $\lambda \in \cC$ has already been dealt with in \cite{Riv1}.
We come back to \eqref{fundeq6.bis} and apply the Bridges-Kennedy-Abdessalam-Rivasseau formula \cite{BK, AR1}. 
We replace the covariance $C_{ij}=1$ by $C_{ij}(x)=x_{ij}$,  $x_{ij}=x_{ji}$ 
(which at the end should be evaluated at $x_{ij}=1$) for $i\neq j$ and $C_{ii}(x)=1$. 

\medskip
Here we start.
\begin{lemma}\label{lemm4}
\bea
\mathfrak{K}^{{\mathcal K}}_{\{abcd\}}&=&
\sum_{T\ \text{LVE tree with ${\mathcal K}$ cilia}}   \cA(T^{\mathcal K}) \, ,\label{eqLVE}\\
\cA(T^{\mathcal K})&=&  (-1)^{|V(T^{\mathcal K})|}
\frac{N^{|V(T^{\mathcal K})|-|E(T^{\mathcal K})|   -{\mathcal K} } }{|V(T^{\mathcal K})|!} \int_0^1 \prod_{e\in E(T)}dt_{e}\, \\
& \times& \int d\mu_{C_{T}}(A) \,  \Tr\Big[\mathop{\overrightarrow{\prod}}\limits_{c\in \partial T\,\text{corner}}\cB_c(i_c)\Big] \, \nonumber\\
&\times&\ \Big[ \frac{\partial^{2}}{\partial J^{+}_{a_{1}b_{1}}\partial J^{}_{c_{1}d_{1}}}\cdots
\frac{\partial^{2}}{\partial J^{+}_{a_{{\mathcal K}}b_{{\mathcal K}}}\partial J^{}_{c_{{\mathcal K}}d_{{\mathcal K}}}}
e^{-\Tr \big(J^{\dagger} \cR^{-1}(x)J\big)} \Big]\bigg|_{\{J\}=0} , \nonumber
\label{eqLVE1}
\ea
where $n = |V(T)|$ is the number of the vertices of the labeled tree,  $v^{T}_{ij} = \inf_{(k,l)\in{ P}_{i\leftrightarrow j}^{T} } t_{kl} $, ${ P}_{i\leftrightarrow j}^{T} $ stands for the unique path joining vertices $i$ and $j$ in the tree $T$, and 
\bea \cR^{-1}(x)&:=&C(x) , \quad  \cR^{-1}_{ij}(x):=x_{ij}.
\ea
\end{lemma}
\prf
Remember  \eqref{eq40}-\eqref{eq41}. In \cite{Saz} it is stated that,
expressing the Gaussian integral as a differential operator \footnote{In \cite{Saz}, Eq. (4.9)-(4.10),
 it is stated for $F(A)$ but we want to apply to $S(A)$.}
\begin{eqnarray}
 \int d\mu_{C(x)}(A)  {\cal S}(A) =
 \left[ e^{\frac{1}{2} \sum_{i,j} x_{ij} \Tr \left[\frac{\partial }{ \partial A_i} \frac{\partial }{ \partial A_j}\right] } {\cal S}(A)\right]_{A_i=0} \, ,
\end{eqnarray}
and
\begin{equation}
\frac{\partial}{\partial x_{ij}}
\bigg(\int d\mu_{C(x)}(A){\cal S}(A)\bigg)=
\frac{1}{2} \int d\mu_{C(x)}(A) \, \Tr \left[\frac{\partial }{ \partial A_i} \frac{\partial }{ \partial A_j}\right] {\cal S}(A) \, .
\end{equation}
The latter differential operator acts on $i$ and $j$ vertices and connects them by an edge. 
The first derivative of the loop vertex for non-polynomial action ${\cal S}(A_i)$ is given by
\bea
&&\hskip-4cm\frac{\partial}{\partial A_{i|cd}} \bigg[
 \Tr\log\left(1-\frac{\myI}{\a}   \frac{\sqrt{\lmbd}}{\sqrt{N}}    A\right)+
\frac{\myI}{\sqrt{N}}\Tr\bigg(A \frac{(1 - \a)}{\sqrt{\lmbd}}\bigg) \bigg] \nonumber\\
\hskip2cm&=& 
-\frac{\myI}{\a}   \frac{\sqrt{\lmbd}}{\sqrt{N}}  \Big(1-\frac{\myI}{\a}     \frac{\sqrt{\lmbd}}{\sqrt{N}}    A_{i}\Big)_{dc}^{-1} +
 \myI \frac{(1 - \a)}{\sqrt{N} \sqrt{\lmbd}  } \mathbb{1}_{dc}\,.
\label{firstD}
\ea
Only the first term of \eqref{firstD} is relevant for applying all further derivatives. Therefore all other derivatives can be computed using the following recursive relation
\begin{equation}
\frac{\partial }{ \partial A_{i|ab}} \frac{-\myI   \sqrt{\lmbd}}{\a\sqrt{N}} \Big(1-\frac{\myI }{\a} \frac{\sqrt{\lmbd}}{\sqrt{N}}A_{i}\Big)_{cd}^{-1}
= -\frac{\lmbd}{\a^2 N} \Big(1-\frac{\myI }{\a}\frac{\sqrt{\lmbd}}{\sqrt{N}}A_{i}\Big)_{ca}^{-1} \Big(1-\frac{\myI }{\a}
\frac{\sqrt{\lmbd}}{\sqrt{N}}A_{i}\Big)_{bd}^{-1}  .
\end{equation}

In the following we attribute factors $\frac{1}{N}$ to the edges of any LVE trees, and define a corner operator 
$\cB$ as\footnote{In \cite{Saz} it is noted $\cC$.} 
\bea\label{eq55}
\cB &=& 
\frac{-\myI\sqrt{\lmbd}}{\a}\Big(1-\text{i}\sqrt{\frac{\lmbd}{\a^2 N}}A_{i}\Big)_{dc}^{-1} + \myI \frac{(1 - \a)}{\sqrt{\lmbd}} \mathbb{1}_{dc}\, 
\ea
if there is one corner in the vertex, and
\bea\label{eq54}
\cB &=& \frac{-\myI\sqrt{\lmbd}}{\a}\Big(1-\text{i}\sqrt{\frac{\lmbd}{\a^2 N}}A_{i}\Big)_{dc}^{-1}\,
\text{     if there are more corners.}
\ea

Now remember Eq. (4.14) of \cite{Saz}. In the present context \emph{with sources $J$} it is stating
\bea 
\mathfrak{K}^{{\mathcal K}}_{\{abcd\}}
&=&  \sum_{T\ \text{LVE tree with ${\mathcal K}$ cilia}}\cA(T^{\mathcal K}) ,
\\ \cA(T^{\mathcal K})&=&(-1)^{|V(T^{\mathcal K})|}\frac{N^{|V(T^{\mathcal K})|-|E(T^{\mathcal K})|}}{|V(T^{\mathcal K})|!} 
\int_0^1 \prod_{e\in E(T{\mathcal K})}dt_{e}\, 
\\ \hskip2cm&\times&  \int d\mu_{C(x)}(A) \Tr\Big[\mathop{\overrightarrow{\prod}}\limits_{c\in \partial T\,\text{corner}}
\cB_c(i_c)\Big] \, \nonumber\\  \nonumber& \times&  \Big[ \frac{\partial^{2}}{\partial J^{+}_{a_{1}b_{1}}\partial J^{}_{c_{1}d_{1}}}\cdots
\frac{\partial^{2}}{\partial J^{+}_{a_{{\mathcal K}}b_{{\mathcal K}}}\partial J^{}_{c_{{\mathcal K}}d_{{\mathcal K}}}}
e^{-N\Tr \big(J^{\dagger} \cR^{-1}(x)J\big)} \Big]\bigg|_{\{J\}=0} . 
\ea
Simply factorise $N$ \emph{in the sources} and we are done.
\qed

\medskip
Remember the equations \eqref{eqLVE}-\eqref{eqLVE1}. Then we decompose \eqref{eqLVE} into five sums: 
\bea
\mathfrak{K}^{{\mathcal K}}_{\{abcd\}}&=& 
\cA(T^{{\mathcal K}}_1) +\sum \cA(T_2^{{\mathcal K}}) 
\nonumber \\&& 
+ \sum_{2 < |V(T^{{\mathcal K}})| < 60} \cA (T^{{\mathcal K}}) + \sum_{T^{{\mathcal K}}_\geq} \cA (T^{{\mathcal K}})
+ \sum_{T^{{\mathcal K}}_<}\cA (T^{{\mathcal K}}),
\label{sumsplit0}
\ea
where, in \eqref{sumsplit0} all the LVE trees have ${\mathcal K}$ cilia, and
\begin{itemize} 
\item the first item is the tree  $T^{{\mathcal K}}_1$,with $\vert V(T^{{\mathcal K}}_1)\vert=1$.  There is only one tree for any ${\mathcal K}$, its corners being equal to the number of cilia. We have pictured in Fig. \ref{T1fig} the case ${\mathcal K}=3$,

\item the second sum is associated to $T_2^{{\mathcal K}}$, see Fig. \ref{T2fig}, always in the case ${\mathcal K}=3$,

\begin{figure}[!ht]\centering\begin{tikzpicture}[>=latex,scale=0.4]
\fill[color=blue,fill=blue!80, very thick](19,0) circle (1); 
\draw[color=green, line width=4pt] (20,0) -- (21,0); \draw[color=green, line width=4pt] (19,1) -- (19,2);
\draw[color=green, line width=4pt] (17,0) -- (18,0); 
\end{tikzpicture}
\caption{Tree $T^3_1$. The cilia are pictured in green.} \label{T1fig}
\end{figure}
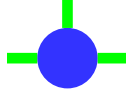

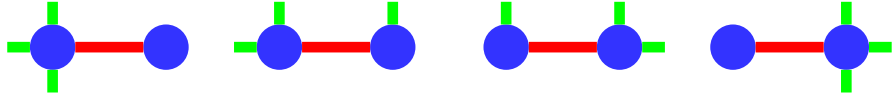
\begin{figure}[!ht]\centering\begin{tikzpicture}[>=latex,scale=0.3]
\fill[color=blue,fill=blue!80, very thick](0,0) circle (1); \fill[color=blue,fill=blue!80, very thick](5,0) circle (1); 
\draw[color=red, line width=4pt] (1,0) -- (4,0);\draw[color=green, line width=4pt] (-2,0) -- (-1,0);`
\draw[color=green, line width=4pt] (0,1) -- (0,2);`\draw[color=green, line width=4pt] (0,-1) -- (0,-2);
\fill[color=blue,fill=blue!80, very thick](10,0) circle (1); \fill[color=blue,fill=blue!80, very thick](15,0) circle (1); 
\draw[color=red, line width=4pt] (11,0) -- (14,0);\draw[color=green, line width=4pt] (8,0) -- (9,0);`
\draw[color=green, line width=4pt] (10,1) -- (10,2);`\draw[color=green, line width=4pt] (15,1) -- (15,2);
\fill[color=blue,fill=blue!80, very thick](20,0) circle (1); \fill[color=blue,fill=blue!80, very thick](25,0) circle (1); 
\draw[color=red, line width=4pt] (21,0) -- (24,0);\draw[color=green, line width=4pt] (25,1) -- (25,2);`
\draw[color=green, line width=4pt] (20,1) -- (20,2);`\draw[color=green, line width=4pt] (26,0) -- (27,0);
\fill[color=blue,fill=blue!80, very thick](30,0) circle (1); \fill[color=blue,fill=blue!80, very thick](35,0) circle (1); 
\draw[color=red, line width=4pt] (31,0) -- (34,0);\draw[color=green, line width=4pt] (35,1) -- (35,2);`
\draw[color=green, line width=4pt] (36,0) -- (37,0);`\draw[color=green, line width=4pt] (35,-1) -- (35,-2);
\end{tikzpicture}
\caption{The amplitudes with $T_2^3$.}
\label{T2fig}
\end{figure}
 
\item the third sum runs over the trees with $2 < |V(T^{{\mathcal K}})| < 60$ number of vertices,
\item  $T^{{\mathcal K}}_\geq$ denotes the LVE trees with ${\mathcal K}$ cilia which have $\al |V(T^{{\mathcal K}})|$ leaves or more and 
$|V(T^{{\mathcal K}})| \geq 60$ vertices, 
\item $T^{{\mathcal K}}_<$ stands for the LVE trees with  vertices $|V(T^{{\mathcal K}})|$ greater or equal to 60
and less than $\al |V(T^{{\mathcal K}})|$ leaves. 
\end{itemize} 

We have to add the sources. We begin by a useful lemma:
\begin{lemma}\label{lemm5}
If $\norm{JJ^+} \le 1$ the equation 
\bea\label{eq63}
\prod_{1\leq m\leq b}\Tr\Big[JJ^{\dagger}
\mathop{\prod}\limits_{1\leq r\leq {\mathcal K}_{m}}^{\longrightarrow}X^{i^{m}_{r}}\Big]=
\sum_{1\leq p_{1}, q_1 \dots \leq N} \prod_{1\leq l\leq {\mathcal K}} X^{l}_{q_{l}p_{\zeta(l)}} \;.
\ea
is true.
\end{lemma}
\prf
Remember \eqref{eq12}. If $\norm{JJ^+} \le 1$, $ (JJ^{\dagger})_{p_{l}q_{l}}   \le 1$. 
Consequently  \eqref{eq63} holds.
\qed

\begin{lemma}
\label{trivialTreeLemma}
Let $1\le {\mathcal K}$ and the condition  $\norm{JJ^+} \le 1$ be fulfilled.
The amplitude of the tree $\cA (T^{{\mathcal K}}_1)$, is analytic in $\lmbd$ and uniformly in $N$ 
on $\C$
with $\lambda \neq 0$ and $-2\pi < \phi < 2\pi$. It is
bounded by 
\begin{eqnarray}
    |\cA (T^{{\mathcal K}}_1)| \leq \frac{N^{1+{\mathcal K}}}{2} \Big\vert\frac{\lambda}{\a^2}\Big\vert \,.
\label{boundAT1}
\end{eqnarray}
\end{lemma}
\prf The lemma is proved in \cite{Saz}, Lemma 3, but \emph{without sources}. 
Since $\norm{JJ^+} \le 1$, we simply add, for any sources, the factor $\vert N^{-{\mathcal K}} \vert = N^{{\mathcal K}}$ and the Lemma is true.
\qed

\begin{lemma}
\label{T2Lemma}
The amplitude of any tree $\cA (T^{{\mathcal K}}_2)$, is analytic in $\lmbd$ and bounded on $\lmbd \neq 0$, $-2\pi < \phi < 2\pi$. 
\end{lemma}
\prf
Using the lemma \ref{boundresolvent}, we obtain
\bea
\Vert\cB\Vert \leq 
\begin{cases}
\Bigg\vert \frac{\sqrt{\lmbd}}{\a} \frac{1}{\cos\psi}\Bigg\vert + \Bigg\vert\frac{(1 - \a)}{\sqrt{\lmbd}}\Bigg\vert\, \text{   if there is only one corner in the vertex,}\\~\\
\Bigg\vert \frac{\sqrt{\lmbd}}{\a} \frac{1}{\cos\psi}\Bigg\vert\, \text{   if there are more corners.}
\end{cases}
\label{cornerbound}
\ea
Remembering the representation \eqref{arep}, for sufficiently large $x$, we can simplify the one-corner bound in \eqref{cornerbound}, requiring that
\begin{eqnarray}
    \Bigg\vert \frac{\sqrt{\lmbd}}{\a} \frac{1}{\cos\psi}\Bigg\vert + \Bigg\vert\frac{(1 - \a)}{\sqrt{\lmbd}}\Bigg\vert \leq  \Bigg\vert\frac{3(1 - \a)}{2\sqrt{\lmbd}}\Bigg\vert\,.
\label{doubleBound}
\end{eqnarray}
It is not hard to see that the latter inequality is valid when
\begin{eqnarray}
    x \geq \max\{x_1, \frac{1}{|\sqrt{\lambda}|}\}\,,\qquad x_1 = \frac{\cos\psi + \sqrt{\cos^2\psi + 8|\lmbd|\cos\psi}}{2 |\sqrt{\lmbd}|\cos\psi}\,.
\label{xineq1}
\end{eqnarray}
Note, as $|\psi| < \pi/2$, we have that $\cos\psi>0$.

Any amplitude $\cA (T^{{\mathcal K}}_2)$ has in all total corners $2+{\mathcal K}$, see Fig. \ref{T2fig}.
Taking into account that there are $2$ vertices, $1=2-1$ edge, and $1+{\mathcal K}$ factors $N$ coming from the trace, add a combinatorial coefficient  related to ${\mathcal K}$, which is $1 + {\mathcal K}=\frac{(1+{\mathcal K})!}{{\mathcal K}!} $ (see Fig. \ref{T2fig}). Therefore we can bound any amplitude $\cA (T^{{\mathcal K}}_2)$  as
\begin{eqnarray}
   \sum \vert\cA(T^{{\mathcal K}}_2)\vert \leq 
\frac{(1+{\mathcal K})!}{{\mathcal K}!} 
   \frac{9 N^{2+{\mathcal K}}}{8} \int d\mu_{C(x)}(A) \Bigg\vert\frac{(1 - \a)}{\sqrt{\lmbd}}\Bigg\vert^2\,.
\end{eqnarray}
Due to the analyticity of the integrand for $|\psi| < \pi/2$ and the latter bound and
since $\norm{JJ^+} \le 1$, we prove Lemma \ref{T2Lemma}.
\qed

\medskip
For the other trees, $\sum_{2 < |V(T^{{\mathcal K}})| < 60} \cA (T^{{\mathcal K}})$, $\sum_{T^{{\mathcal K}}_\geq} \cA (T^{{\mathcal K}})$ and
$\sum_{T^{{\mathcal K}}_<}\cA (T^{{\mathcal K}})$ remember \cite{Saz}, Lemma 5 to Lemma 10. Adding sources is very easy if we take into account Lemma \ref{lemm5}, but nevertheless we want to state the analog of these lemmas.

We establish the bound for the corner operators of a general tree. But let us start by an other useful lemma

\begin{lemma}
\label{lemma8}
Provided  $\norm{JJ^+} \le 1$ be fulfilled, the following equation holds: 
\bea 
&&\hskip-2cm \Big[ \frac{\partial^{2}}{\partial J^{+}_{a_{1}b_{1}}\partial J^{}_{c_{1}d_{1}}}\cdots
\frac{\partial^{2}}{\partial J^{+}_{a_{{\mathcal K}}b_{{\mathcal K}}}\partial J^{}_{c_{{\mathcal K}}d_{{\mathcal K}}}}
e^{-\Tr \big(J^{\dagger} \cR^{-1}(x)J\big)} \Big]\bigg|_{\{J\}=0}\nonumber
\\&& \hskip2cm =
\bigg[ \prod_{l=1}^{3({\mathcal K}-1)}\cD_{m_{l+1} m_{l+2}}\bigg] \cD_{m_{3{\mathcal K}+1} m_{3{\mathcal K}+2}} 
\label{eq70}
\ea
\end{lemma}
\prf If we remember Proposition \ref{structure:prop}, in particular \eqref{structure:eq}
and \eqref{structure:eq1}, and Lemma \ref{lemm5}, one could deduce easily that \eqref{eq70} holds.
\qed

\begin{figure}[!ht]\centering\begin{tikzpicture}[>=latex,scale=0.5]
\fill[color=black,fill=blue!80, very thick](9,-4) circle (1) ;\fill[color=black,fill=blue!80, very thick](13,-4) circle (1);\draw[color=red, line width=4pt] (10,-4) -- (12,-4);
\fill[color=black,fill=blue!80, very thick](13,-8) circle (1); \draw[color=red, line width=4pt] (13,-5) -- (13,-7);\fill[color=black,fill=blue!30, very thick](12,0) circle (1); \draw[color=red, line width=4pt] (17.5,-3.3) -- (18.5,-1.7);
\draw[->,dashed,color=red, line width=4pt] (12.3,-1) -- (12.8,-3);\draw[->,dashed,color=red, line width=4pt] (16.3,-1) -- (16.8,-3);\fill[color=black] (14.5,-4) circle (-0.2) ;
\fill[color=black] (15,-4) circle (-0.2) ;\fill[color=black] (15.5,-4) circle (-0.2) ; \fill[color=black,fill=blue!80, very thick](17,-4) circle (1); \fill[color=black,fill=blue!30, very thick](16,0) circle (1); \fill[color=black,fill=blue!80, very thick](19,-1) circle (1) ;\fill[color=black,fill=blue!80, very thick](21,-4) circle (1); 
\draw[color=red, line width=4pt] (18,-4) -- (20,-4);\fill[color=black,fill=blue!30, very thick](25,-4) circle (1); \draw[<-,dashed,color=red, line width=4pt] (22,-4) -- (24,-4);\fill[color=black,fill=blue!80, very thick](21,-8) circle (1); 
\draw[color=red, line width=4pt] (21,-5) -- (21,-7);\end{tikzpicture}\caption{Possible ways to add a leaf to a tree.}\label{proof2}
\end{figure}
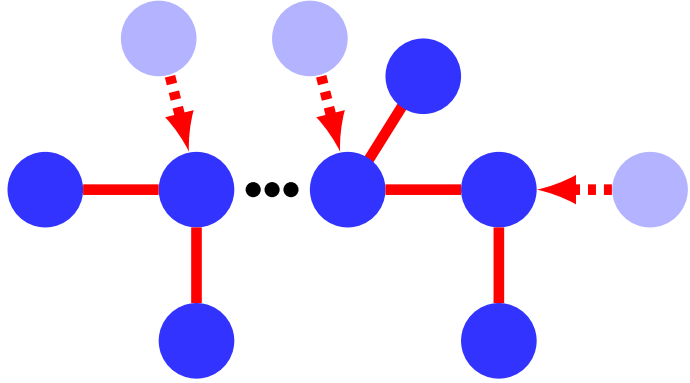

\medskip
Let us state  the combinatorial lemmas of \cite{Saz}, this time with ${\mathcal K}$, hence with the sources.
But to illustrate these lemmas, we reproduce two figures taken from \cite{Saz}, Figures \ref{treesFig} and \ref{proof2}.

\begin{figure}[!ht]\centering\begin{tikzpicture}[>=latex,scale=0.5]\draw (-1,4) node{$1)$} ;
\fill[color=black,fill=blue!80, very thick](0,0) circle (1) ;\fill[color=black,fill=blue!80, very thick](3,3) circle (1) ;\draw[color=red, line width=4pt] (0.7,0.7) -- (2.3,2.3);\fill[color=black,fill=blue!80, very thick](4,0) circle (1); \draw[color=red, line width=4pt] (1,0) -- (3,0);\fill[color=black] (2.5,-1) circle (-0.2) ;\fill[color=black] (2.2,-1.5) circle (-0.2) ;\fill[color=black] (1.5,-2) circle (-0.2) ;\fill[color=black,fill=blue!80, very thick](0,-4) circle (1);
\draw[color=red, line width=4pt] (0,-1) -- (0,-3); \draw (7,4) node{$2)$} ;\fill[color=black,fill=blue!80, very thick](9,3) circle (1) ;\fill[color=black,fill=blue!80, very thick](13,3) circle (1);\draw[color=red, line width=4pt] (10,3) -- (12,3);\fill[color=black] (14.5,3) circle (-0.2) ;\fill[color=black] (15,3) circle (-0.2) ;\fill[color=black] (15.5,3) circle (-0.2) ; \fill[color=black,fill=blue!80, very thick](17,3) circle (1); \fill[color=black,fill=blue!80, very thick](21,3) circle (1); 
\draw[color=red, line width=4pt] (18,3) -- (20,3);\draw (7,-4) node{$3)$} ;\fill[color=black,fill=blue!80, very thick](9,-4) circle (1) ;\fill[color=black,fill=blue!80, very thick](13,-4) circle (1);\draw[color=red, line width=4pt] (10,-4) -- (12,-4);\fill[color=black,fill=blue!80, very thick](13,-8) circle (1); \draw[color=red, line width=4pt] (13,-5) -- (13,-7);\draw[color=red, line width=4pt] (17.5,-3.3) -- (18.5,-1.7);
\fill[color=black] (14.5,-4) circle (-0.2) ;\fill[color=black] (15,-4) circle (-0.2) ;\fill[color=black] (15.5,-4) circle (-0.2) ; \fill[color=black,fill=blue!80, very thick](17,-4) circle (1); \fill[color=black,fill=blue!80, very thick](19,-1) circle (1) ;
\fill[color=black,fill=blue!80, very thick](21,-4) circle (1); \draw[color=red, line width=4pt] (18,-4) -- (20,-4);
\fill[color=black,fill=blue!80, very thick](21,-8) circle (1); \draw[color=red, line width=4pt] (21,-5) -- (21,-7);
\end{tikzpicture}\caption{1) A tree with a maximal amount of leaves at the given order of the loop vertex expansion. 2) A tree with a minimal amount of leaves at the given order of the LVE. 3) Representation of an average LVE tree with not so many leaves.}\label{treesFig}
\end{figure}
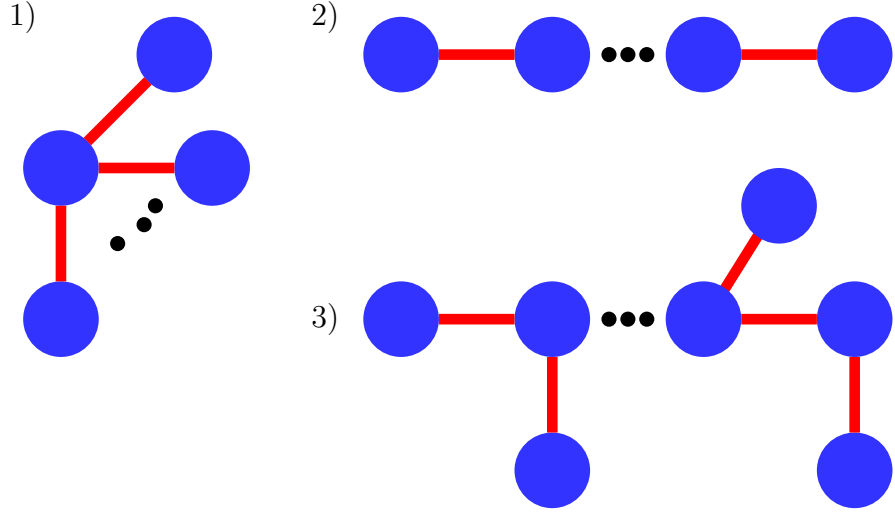

\begin{lemma}\label{cornerLemmabis}
If $\Vert JJ^\dagger\Vert\le 1 $, if a tree $T^{\mathcal K}$ has $n_l$ leaves, $n_i$ internal vertices and ${\mathcal K}$ cilia, and if $n_l + n_i = |V(T^{\mathcal K})| > 2$, 
its  trace of the product of the corner operators is bounded by
\bea
\bigg| \Tr\Big[\mathop{\overrightarrow{\prod}}\limits_{c\,\in\, \partial T_{(n_i, n_l)}\,\text{corner}}
\cB_c(i_c)\Big]\bigg| &\le&  \frac{(n+{\mathcal K}-1)!}{(n-1)!{\mathcal K}!}.N^{1+{\mathcal K}}  \nonumber
\\&&\Bigg\vert \frac{\sqrt{\lmbd}}{\a} \frac{1}{\cos\psi}\Bigg\vert^{2 n_i - 2} \Bigg\vert\frac{3(1 - \a)}{2 \a \cos\psi}\Bigg\vert^{n_l}\, .
\label{eqTreeABound2bis}
\ea 
\end{lemma}
\prf
Remember Lemma 5 of \cite{Saz}, and Lemmas \ref{lemm4} and \ref{lemm5} of this paper. Then one could deduce easily that 
\eqref{eqTreeABound2bis} holds.
\qed

\begin{lemma}
\label{allbyConstLemmabis}
For any $x$ satisfying
\bea
x \geq x_2\,,\qquad x_2 &=&\frac{|\sqrt{\lmbd}| + 1}{|\sqrt{\lmbd}|}\,,
\label{x2condbis}
\ea
one has
\bea
&&\Bigg\vert \frac{\sqrt{\lmbd}}{\a} \frac{1}{\cos\psi}\Bigg\vert \leq \Bigg\vert\frac{(1 - \a)}{\a \cos\psi} \Bigg\vert \leq \frac{\sqrt{2}}{\cos\psi} \,.
\label{twoIneq0bis}
\ea
\end{lemma}
\prf
Simply quote Lemma 6 of \cite{Saz}, and 
\eqref{twoIneq0bis} holds.
\qed

\begin{lemma}
\label{FiniteSumLemma}
For $|\psi| < \pi/2$, $\lambda \neq 0$ the sum of absolute values of trees amplitudes from the finite set of trees with ${\mathcal K}$ cilia in \eqref{sumsplit0} 
is bounded by a constant,
\bea\label{eq73}
\sum_{2 < |V(T^{\mathcal K})| < 60} |\cA_{T^{\mathcal K}}| < const\,,
\ea
where the $const\, \sim \, O(N^{2+{\mathcal K}})$.
\end{lemma}
\prf
Remember Lemma 7 of \cite{Saz}, and Lemmas \ref{lemm4} and \ref{lemm5} of this paper. Then one could deduce easily that 
\eqref{eq73} holds.
\qed

\begin{lemma}
\label{LeavesLemma}
The number of trees with ${\mathcal K}$ cilia, with $|V(T^{\mathcal K})|$ vertices and 
$\al |V(T^{\mathcal K})|$ or more leaves with $1/2 < \al < 1$, 
$|T^{\mathcal K}_\geq|$, is bounded by
\bea
    |T^{\mathcal K}_\geq| &\leq&  
2^{3n+{\mathcal K}}\ (n-1)!    (|V(T^{\mathcal K})| - \ceil*{\al |V(T^{\mathcal K})|}  + 1)
    \big(|V(T^{\mathcal K})|!\big) \nonumber\\
    &\times& 2^{|V(T^{\mathcal K})|-3} e^{\ceil*{\al |V(T^{\mathcal K})|}} \bigg(\frac{1 - \al}{\al}\bigg)^{\ceil*{\al |V(T^{\mathcal K})|}}\,.
\label{eq74}
\ea
\end{lemma}
\prf
Remember Lemma 8 of \cite{Saz}, and Lemmas \ref{lemm1}, \ref{lemm4} and \ref{lemm5} of this paper. Then one could deduce easily that 
\eqref{eq74} holds.
\qed

\begin{lemma}
\label{TgeqLemma}
For $|\psi| < \pi/4$ and $\al = \frac{59}{60}$, the sum of absolute values of amplitudes of the trees $T^{\mathcal K}_\geq$ is smaller than the sum of an absolutely convergent series,
\bea\label{eq75}
\sum_{T^{\mathcal K}_\geq} \big\vert \cA_{T^{\mathcal K}_\geq}\big \vert \leq 
2^{3n+{\mathcal K}}\  (n-1)!     \frac{N^{2+{\mathcal K}} e}{8}\sum_{v=60}^\infty (\frac{1}{60} v  + 1)\, \Bigg(\frac{7}{8}\Bigg)^{v}\,.
\ea
\end{lemma}
\prf
Remember Lemma 9 of \cite{Saz}, and Lemmas \ref{lemm1}, \ref{lemm4} and \ref{lemm5} of this paper. Then one could deduce easily that 
\eqref{eq75} holds.
\qed

\begin{lemma}
\label{TlessLemma}
For $|\psi| < \pi/4$ and $\al = \frac{59}{60}$, the sum of absolute values of amplitudes of the trees $T^{\mathcal K}_<$ is smaller than the sum of an absolutely convergent series
\bea
\sum_{T^{\mathcal K}_<} \big\vert \cA_{T^{\mathcal K}_<}\big \vert &\leq&
2^{3n+{\mathcal K}}\  (n-1)!     N^{2+{\mathcal K}} \  \frac{2 (x_4 \cos\psi)^5 \sqrt{\lmbd}}{3 (x_4\sqrt{\lmbd} - 1)}\sum_{v=60}^\infty \frac{1}{v^2}\Bigg(\frac{1}{2}\Bigg)^{v}\,,
\label{eq76}
\ea
where
\bea
x_4 = \max\{x_3, \frac{2^{30}e^{30}}{\cos\psi} \Big(\frac{3\sqrt{2}}{2\cos\psi}\Big)^{59/2}\} + 1\,.
\label{eq77}
\ea
\end{lemma}
\prf
Remember Lemma 10 of \cite{Saz}, and Lemmas \ref{lemm1}, \ref{lemm4} and \ref{lemm5} of this paper. 
Then one could deduce that \eqref{eq76} and \eqref{eq77} holds.
\qed

\section{Proof of the two results}

\subsection{Proof of Theorem \ref{mainTheorem1}}
\medskip
In lemmas \ref{trivialTreeLemma}, \ref{T2Lemma}, \ref{FiniteSumLemma}, \ref{TgeqLemma}, and \ref{TlessLemma} we proved absolute and uniform in $N$ convergence of the series \eqref{eqLVE}. The terms of \eqref{eqLVE} are analytic functions of $\lambda$, if $\lambda \neq 0$, for any fixed values of $x \geq x_4$ and $|\psi| < \frac{\pi}{4}$ from the representation of the variational parameter $\a = x \sqrt{\lambda} e^{i\psi}$. Let us now show that the value $x \geq x_4$ can be chosen independently of $\lambda$. Recalling definitions of $x_1$, $x_2$, $x_3$, we can rewrite $x_4$ as
\bea
x_4 = &&\max\Bigg\{\frac{\cos\psi + \sqrt{\cos^2\psi + 8 |\lambda|\cos\psi}}{2 |\sqrt{\lambda}|\cos\psi}, \nonumber\\
&&\hskip2cm\frac{|\sqrt{\lambda}| + 1}{|\sqrt{\lambda}|}, \frac{2^{30}e^{30}}{\cos\psi} \Big(\frac{3\sqrt{2}}{2\cos\psi}\Big)^{59/2}\Bigg\} + 1\,.
\ea
Obviously, $x_4$ is bounded for $|\lambda| > 0$ and $|\psi| < \pi/4$. Therefore, it is enough to consider the variational parameter $a$ with the fixed value of $x = x_5$, where
\begin{equation}
    x_5 = \max_{|\lambda|, \psi} x_4\,.
\end{equation}

\medskip
Taking $\psi < \frac{\pi}{4}-\frac{\epsilon}{2}$ leads to the convergence of \eqref{eqLVE} and to the analyticity of its terms for $-\frac{\pi}{2} + 2\psi < \phi < \frac{\pi}{2} + 2\psi$, $|\lambda| > 0$.
Thus we obtain the analytic continuation of the  cumulants $\mathfrak{K}^{{\mathcal K}}_{\{abcd\}}$ for $|\phi| < \pi - \epsilon$, $|\lambda| > 0$, and  for any $ \epsilon$, $0<\epsilon\le \frac{\pi}{2}$.

\subsection{Proof of Theorem \ref{mainTheorem2}}

\medskip
We define ${\mathfrak{K}}^{\mathcal K}_{\pi}(\lambda, N)$ as in \eqref{eq46}.
Recall Proposition 2, Theorem 1, Theorem 2, Corollary 1, Theorem 3, Lemma 1 and Lemma 2 common to \cite{GurKra} and to this paper.

We have to prove the part associated to $\lambda \in {\cal E}$. To do this, we employ the same strategy than to deal with the case of Theorem \ref{mainTheorem1}; we proved absolute and uniform in $N$ convergence of the series \eqref{eq46}  associated to $\lambda \in {\cal E}$
by reminding the  lemmas \ref{trivialTreeLemma}, \ref{T2Lemma}, \ref{FiniteSumLemma}, \ref{TgeqLemma}, and \ref{TlessLemma} and the 
content of the previous subsection.
 
\medskip
The rest of the theorem \ref{mainTheorem2}, in particular \eqref{treecumulantsbound1}, is trivial, once we know \cite{GurKra}. 

\section{Appendix : Nevanlinna-Sokal theorem}
\label{sect8}
We recall the following theorem: 

\begin{figure}[!ht]
\begin{center}\begin{tikzpicture}[scale=1.2]
\fill[color=red!40] (1,0) circle (1) ;
\draw[->] (-1,0) -- (2.5,0);
\draw [->] (0,-1.5) -- (0,1.5);
\draw[red] (1,0) circle (1) ;
\draw[red] (1.1,.7) node[below]{$^{R}$} ;
\draw[red] (1.5,-1.5) node[below]{$\cD_R$} ;
\draw[red,<->] (1,0) -- (1.7071,0.7071);
\draw[red] (1,0) circle (1) ;
\fill[color=red!40] (5,1) arc (90:270:1) -- (9,-1) -- (9,1) -- cycle ;
\draw[->] (3.5,0) -- (9.5,0);
\draw [->] (5,-1.5) -- (5,1.5);
\draw[red] (5,1) arc (90:270:1) -- (9,-1) -- (9,1) -- cycle ;
\draw[red,<->] (5,0) -- (4.2929,0.7071);
\draw[red] (4.4,0.5) node[below]{$^{\sigma^{-1}}$} ;
\draw[red] (6,-1.5) node[below]{$\Sigma_\sigma$} ;
\end{tikzpicture}
\end{center}
\caption{Domain of analyticity of $F$ and of its Borel transform for $q=1$.}
\label{CuX}
\end{figure}
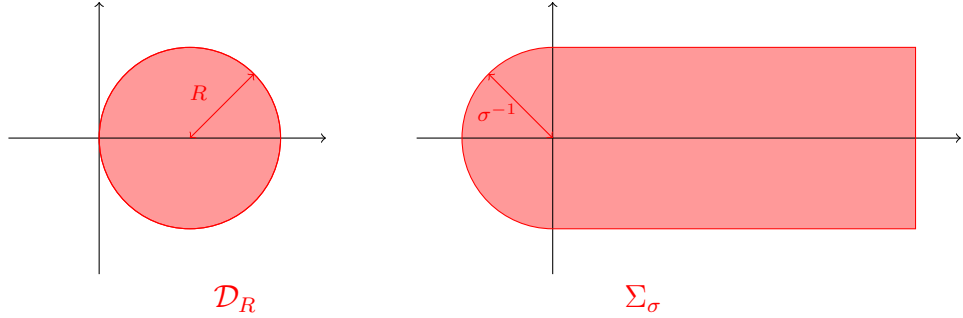

\begin{theorem}[Nevanlinna-Sokal \cite{Nev,Sokal}]
\label{NevanlinnaSokal} Let $R>0$ and $F_{\omega}(\lambda)$ be a family of analytic functions on the disc ${\cal D}_R$ depending on some 
parameter $\omega\in\Omega$. If there exists a sequence $a_{n}(\omega)$ of  functions of $\omega\in\Omega$ obeying, 
for any $n$, $\lambda\in{\cal D}_{R}$ and $\omega\in\Omega$ the uniform bound: 
  \begin{equation}
 \big| F_{\omega}(\lambda)-\sum_{m=0}^{n}a_{m}(\omega)\lambda^{m} \big|< C\sigma^{n+1}|\lambda|^{n+1}(n+1)!\label{Taylorbound} \; ,
 \end{equation}
 with $C$ and $\sigma$ two positive constants that do not depend on $\omega$, then the series
 \begin{equation}
B_{\omega}(s)=\sum_{n=0}^{\infty}\frac{a_{n}(\omega)}{n!}s^{n} \;,
 \end{equation}
 has radius of convergence $\sigma^{-1}$ and can be analytically continued in the strip $\Sigma_{\sigma}$.
Moreover, there exists a constant $B$ such that, for any $s\in\Sigma_{\sigma}$ and $\omega\in\Omega$, we have
\begin{equation}
\big|B_{\omega}(s)\big|\leq B \text{e}^{\frac{s}{R}} \;.
\end{equation}
Finally, for any $\lambda\in{\cal D}_{R}$, $F_{\omega}(\lambda)$ is given by the following absolutely convergent integral:
\begin{equation}
F_{\omega}(\lambda)=\int_{0}^{\infty}\!\!ds \,B_{\omega}(s)
\text{e}^{-\frac{s}{\lambda}}\label{Boreltransform} \; .
\end{equation}
\end{theorem}

\begin{remark}
This Theorem is the corner stone of our group \cite{Rivbook,GurKra,Guraubook}. It has been stated with absolute clarity and it includes Borel-LeRoy formulation in \cite{CaGrMa}.
\end{remark}

\end{document}